\documentclass[preprint]{revtex4-1}
\usepackage[english]{babel}
\usepackage{amsmath}
\usepackage{amssymb}
\usepackage[final]{graphicx}
\usepackage{upgreek}
\usepackage[utf8]{inputenc}
\usepackage{latexsym}

\begin{document}
\title{Coexistence of strong and weak coupling in ZnO nanowire cavities}
\author {Tom Michalsky}
\affiliation{Institut f\"{u}r Experimentelle Physik II, Universit\"{a}t Leipzig, Linn\'{e}stra\ss e 5, D-04103 Leipzig}

\author{Helena Franke}
\affiliation{Institut f\"{u}r Experimentelle Physik II, Universit\"{a}t Leipzig, Linn\'{e}stra\ss e 5, D-04103 Leipzig}

\author{Robert Buschlinger}
\affiliation{Institut f\"{u}r Festk\"{o}rpertheorie und -optik, Friedrich-Schiller-Universit\"{a}t Jena, Max-Wien-Platz 1, 07743 Jena, Germany}

\author{Ulf Peschel}
\affiliation{Institut f\"{u}r Festk\"{o}rpertheorie und -optik, Friedrich-Schiller-Universit\"{a}t Jena, Max-Wien-Platz 1, 07743 Jena, Germany}

\author{Marius Grundmann}
\affiliation{Institut f\"{u}r Experimentelle Physik II, Universit\"{a}t Leipzig, Linn\'{e}stra\ss e 5, D-04103 Leipzig}

\author{R\"{u}diger~Schmidt-Grund}
\affiliation{Institut f\"{u}r Experimentelle Physik II, Universit\"{a}t Leipzig, Linn\'{e}stra\ss e 5, D-04103 Leipzig}

\begin{abstract}
We present a high quality two-dimensional cavity structure based on ZnO nanowires coated with concentrical Bragg reflectors. The spatial mode distribution leads to the simultaneous appearance of the weak and strong coupling regime even at room temperature. Photoluminescence measurements agree with FDTD simulations. Furthermore the ZnO core nanowires allow for  the observation of middle polariton branches between the A- and B-exciton ground state resonances. Further, lasing emission up to room temperature is detected in excitation dependent photoluminescence measurements.   
\end{abstract}
%
%
\maketitle
\section{Introduction}
\label{sec:intro}

Light-matter interaction plays a key role in the development of integrated optoelectronic devices. Such devices could be based on properties of exciton-polaritons. These bosonic quasi-particles are able to undergo Bose-Einstein condensation (BEC), show superfluidity and pa\-ra\-me\-tric oscillations~\cite{Kasprzak2006,Amo2009,Kundermann2003}. Furthermore the complex optical mode structure in two-dimensionally confined wire cavities can be used to realize parametric interbranch scattering~\cite{Dietrich2015} which should enable the emission of entangled photons~\cite{Ciuti2004}. Therefore nanowire cavity based devices promise to be a versatile platform for sources of entangled photons. Their linear geometry could also play a major role in the realization of integrated guides for superfluid condensates.

In literature dealing with with wire-like cavities often two mode families are discussed. One type are the so-called whispering gallery modes (WGMs) which are confined due to total internal reflection (TIR) at the cavity-ambient interface~\cite{Wiersig2003,Czekalla2008,Markushev2012,Grundmann2012}. These modes in general feature high quality factors as minimal losses are connected with TIR. Here all the cavity losses, apart from material absorption, are connected to the cavity corners (in a hexagonal cavity for example) or surface roughnesses~\cite{Wiersig2003}. Typical high quality WGM cavities have a diameter much larger than the light wavelength of the confined photons. This leads to a spatial field distribution of WGMs only in the outermost region of the cavity. The other type are Fabry-P\'{e}rot modes (FPM) which are formed through multiple reflections between plane parallel interfaces~\cite{Dietrich2011,Coulon2012} at incident angles below the critical angle for TIR. An adequate pendant in a circular cavity are Bessel modes~\cite{Yeh1978}. Nevertheless, similar to FPMs in planar cavities, these modes lack high quality factors in bulk cavity structures as the reflectivity at the cavity-ambient interface is always well below $100\ \%$. Therefore planar cavities are coated with highly reflecting distributed Bragg reflectors (DBRs) to increase the quality factor of FPMs. 

Within our approach we use self-assembled grown ZnO nanowires as active cavity material. To achieve a strong lateral photonic confinement these nanowire cavities are coated concentrically with DBRs~\cite{RSG10_PSS}. In this paper we show that two-dimensionally quantized modes arise in these cavities showing different coupling regimes to the ZnO excitons namely the weak (WCR) and the strong coupling regime (SCR)~\cite{Savona1995}. These modes are able to propagate along the cavity axis. The considerably high quality of our ZnO core material and the multimode character of the 2D cavity enables the observation of multimode exciton-polaritons which simultaneously couple to the lowest excitonic states yielding multiple lower and middle polariton branches at low temperatures. Our experimental findings are supported by finite difference time domain (FDTD) simulations. Further we observe nonlinear emission effects up to room temperature arising in the weakly coupled modes. We propose that this is originated from electron-hole plasma induced lasing.  


\section{Sample preparation}
\label{sec:prep}

The nanowire cavities are produced by pulsed laser deposition (PLD) in three steps. First a ZnO nucleation layer is deposited on a sapphire substrate. The nucleation layer thickness determines the density of the nanowires which are grown in the second PLD step under low vacuum conditions \cite{Cao2009}. The nanowire density is adjusted such that no shadowing occurs during the deposition of the DBR layers with oblique incidence PLD. The DBR of the nanowire cavity discussed within this paper consists of 10.5 layer pairs YSZ and $\mathrm{Al}_2\mathrm{O}_3$ being concentrically deposited on the ZnO nanowire.
More details regarding the growth process can be found in Ref.~\cite{RSG10_PSS}.    

\begin{figure*}
\centering 
\resizebox{1\textwidth}{!}{%
  \includegraphics{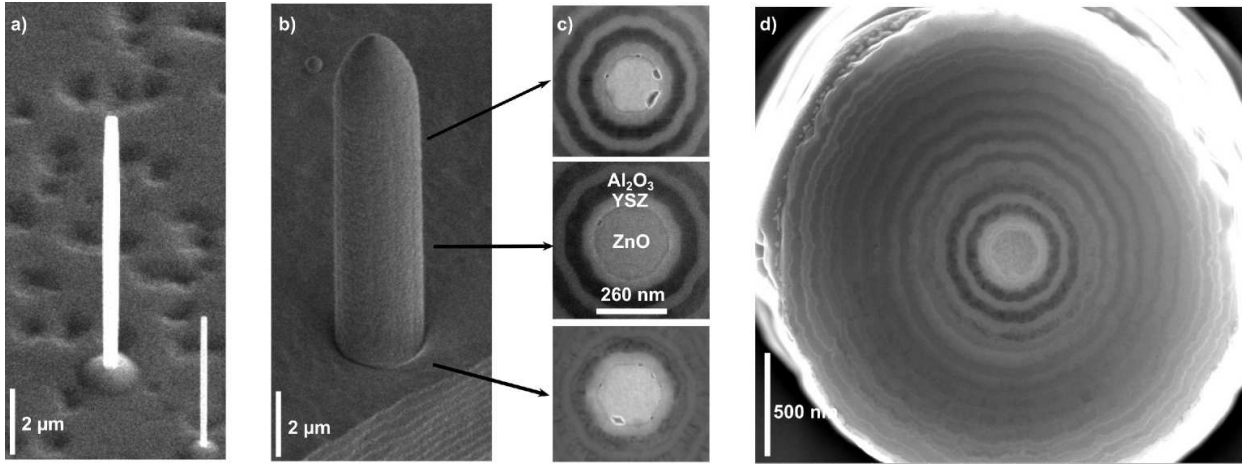}}
\caption{Scanning electron microscopy image of (a) a bare ZnO nanowire and (b) the whole cavity structure . The images in (c) show the center of the cavity and have been taken after cutting the cavity into slices with a focused ion beam.(d) Cross-section of the whole cavity structure. The rough surface at the DBR-air interface is introduced by an extra platinum layer which was deposited before cutting with the focused ion beam to avoid charge accumulation.}
\label{fig:sem}    
\end{figure*}

The cavity being investigated within this paper has a ZnO core with a diameter of about $\sim 260\ \mathrm{nm}$ in the middle of the cavity's length axis. Towards both ends the wire is slightly tapered as can be seen from Fig.~\ref{fig:sem}. Furthermore at the center the core wire has an almost perfect circular geometry, whereas towards the ends the wire geometry changes to be more hexagonal. In order to obtain this information on geometry the cavity was cut into slices with a focused ion beam (FIB) and investigated with a scanning electron microscope after all other characterizations had been performed. Irregularities at the core wire-DBR interface can be seen in Fig.~\ref{fig:sem} c). These may result from the FIB cutting itself where material is pulled out at different rates. 

\section{Experimental and computational details}
\label{sec:experiment}
The optical investigations were performed with a micro imaging setup which is capable of imaging spatially resolved the angular as well as the real space emission or reflection from a sample surface on a monochromator entrance slit. We used a Mitutoyo UV objective with a magnification of 50 and a numerical aperture (NA) of 0.4 corresponding to a detectable angular range of $\pm 23^\circ$. For confocal photoluminescence (PL) linescans a pinhole was put near the backfocal plane of the objective to select a certain angular range while keeping a high spatial resolution by reading out only one CCD row.
In case of angular resolved measurements the spatial resolution was limited by the excitation spot size of about $2\ \mathrm{\upmu m}$ whereas for confocal linescans a spatial resolution of about  $1\ \mathrm{\upmu m}$ was set.

PL excitation was performed either with a continuous wave (Coherent Verdi V10 + MDB 266) or a pulsed (frequency doubled Coherent Vector 532) laser source  with a wavelength of $266\ \mathrm{nm}$ for nonresonant excitation.
For temperature dependent measurements the sample was put in a helium flow cryostat (JANIS ST-500). The customized sample holder inside the cryostat allowed for an oblique angle between the wire and the optical axis of the imaging system resulting in an observable angular range from $-4^\circ$  to $42^\circ$ with respect to the normal of the wire axis.

For the FDTD simulations, we use the commercial software Lumerical FDTD Solutions. The energy-dependent refractive indices of the cavity materials used in the simulations are obtained by a model analysis of experimental data as shown in the Appendix.
In order to extract eigenmodes and their energies using the FDTD method, we simulate illumination of the structure with a spectrally broad signal exciting multiple modes. The excitation has the spatial shape of a plane wave, since we are, for comparison with the experiment, only interested in cavity-modes which can be excited by a propagating plane-wave. After the initial pulse has passed the structure, eigenenergies are indicated by the peaks in the  fourier transform of the time evolution of the fields inside the cavity.

\section{FDTD simulations}
\label{sec:fdtd}

\begin{figure*}
\centering 
\resizebox{1\textwidth}{!}{%
  \includegraphics{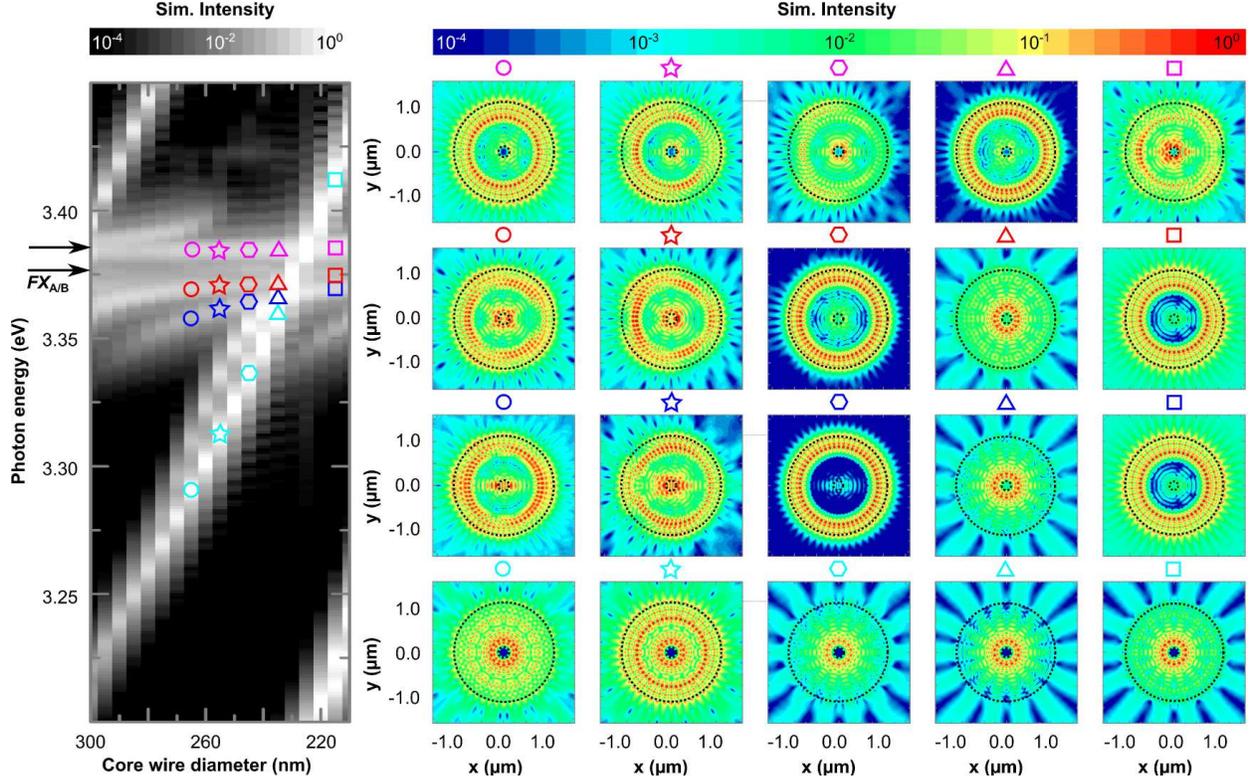}}
\caption{FDTD simulations: the left panel shows the spectrally resolved field intesities for polarization perpendicular to the wire axis for different ZnO core wire diameters. The right panel shows an image array of the spatial intensity distributions for a decreasing core wire diameter (left to right) and for the different modes (bottom to top). The colored symbols indicate the corresponding energy and core wire diameter in the left panel. In the spatial intensity images the core wire-DBR and the DBR-air interfaces are indicated as black dotted circles.} 
\label{fig:simulation}    
\end{figure*}

In order to gain insight in the spatial intensity distribution of resonant modes inside our cavity structure, which is \textit{a priori} not possible to conclude from optical measurements alone, we performed two-dimensional FDTD simulations. Here, only fields propagating in the plane perpendicular to the wire axis were investigated.
The detuning between photonic and excitonic modes, which gives access to their underlying coupling mechanisms, was thereby varied by changing the core wire diameter ($\sim$ the cavity thickness) in the range from $210\ \mathrm{nm}$ to  $300\ \mathrm{nm}$. We have chosen a radially symmetric geometry of the ZnO core wire as well as of the DBR as our cavity exhibits this geometry in its center (see Fig.~\ref{fig:sem}).
In the left panel of Fig.~\ref{fig:simulation} the simulated spectral field intensities for polarization perpendicular to the wire axis (TE) are shown as a function of the core wire diameter. Furthermore some selected field distributions of resonant modes are shown in Fig.~\ref{fig:simulation} (right). From the simulated spectra one can clearly identify modes with different dispersions $E(d)$ with regard to the core wire diameter $d$.

The brightest mode, having an energy of $3.21\ \mathrm{eV}$ at $300\ \mathrm{nm}$ core wire thickness, clearly crosses the excitonic A resonance at $FX_{\mathrm{A}}=3.376\ \mathrm{eV}$ for a core wire diameter of $225\ \mathrm{nm}$ and the excitonic B resonance at $FX_{\mathrm{B}}=3.385\ \mathrm{eV}$ for a core wire diameter of $220\ \mathrm{nm}$ (light blue symbols in Fig.~\ref{fig:simulation}). By following the evolution of the spatial field distribution with diameter it becomes clear that this mode has always a vanishing field amplitude inside the ZnO core wire. This means that there is no spatial overlap between the cavity photons and the ZnO excitons, which results in a weakly or decoupled subsystem. This is reflected by the observed mode crossing behaviour \cite{Savona1995}. 

In contrast to that we find modes showing a clear anticrossing behaviour by approaching the A and B excitonic resonances. These modes are labeled with red and blue symbols in Fig.~\ref{fig:simulation}.
With decreasing core wire diameter their mode energies asymptotically converge to the energy of the A-exciton at  $3.376\ \mathrm{eV}$. In contrast to the mode mentioned before, one always finds field intensity inside the core wire. This means that there is a nonvanishing mode overlap between the cavity photons and the core wire excitons enabling the strong coupling for these modes. In general such strongly coupled modes at energies below the excitonic transitions are called lower polariton branches.   

There also exists a mode showing no dispersion with changing cavity thickness. That mode is centered at $3.38\ \mathrm{eV}$ in the spectral region right below the B-exciton resonance. 
The corresponding spatial field distributions (pink symbols in Fig.~\ref{fig:simulation}) are not that regularly as in the case of the modes mentioned before. But also here we can find a field amplitude inside the core wire. As we will show later, we attribute this mode to be the sum of several middle polariton branches. 

Furthermore an upper polariton branch appears in the FDTD simulated spectra. It is specified through its anticrossing behaviour when approaching the excitonic B resonance at $3.385\ \mathrm{eV}$ from the high energy spectral range with increasing core wire diameter. Also for this mode a nonvanishing field intensity in the core wire region is found (not shown).

All modes exhibit a spectral width of $\gamma= 2.5\pm 0.5\ \mathrm{meV}$ (half width at half maximum-HWHM) near the excitonic resonance resulting in a quality factor of $Q=E/(2\gamma)$ of the cavity of about 700.     

In general we conclude from the FDTD simulations that neither pure WG- nor FP-like modes form in our cavity structure. We always find modes having many nodes in the azimuthal direction, which is indeed typical for WGMs, but most of the modes are also widely distributed inside the entire cavity cross section including the DBR. We do not find any modes having their maximum field intensity located at the center of the cavity structure and showing an exponential decay of the field inside the DBR which would be typical for FP like modes. 
The main criteria to distinguish different modes is if they have spatial overlap with the core wire or not. This directly leads to the separation in modes being in the SCR or WCR with the typical anticrossing or crossing behaviour.

\section{Experimental results}
\subsection{Photonic confinement}
\begin{figure*}
\centering 
\resizebox{1\textwidth}{!}{%
 \includegraphics{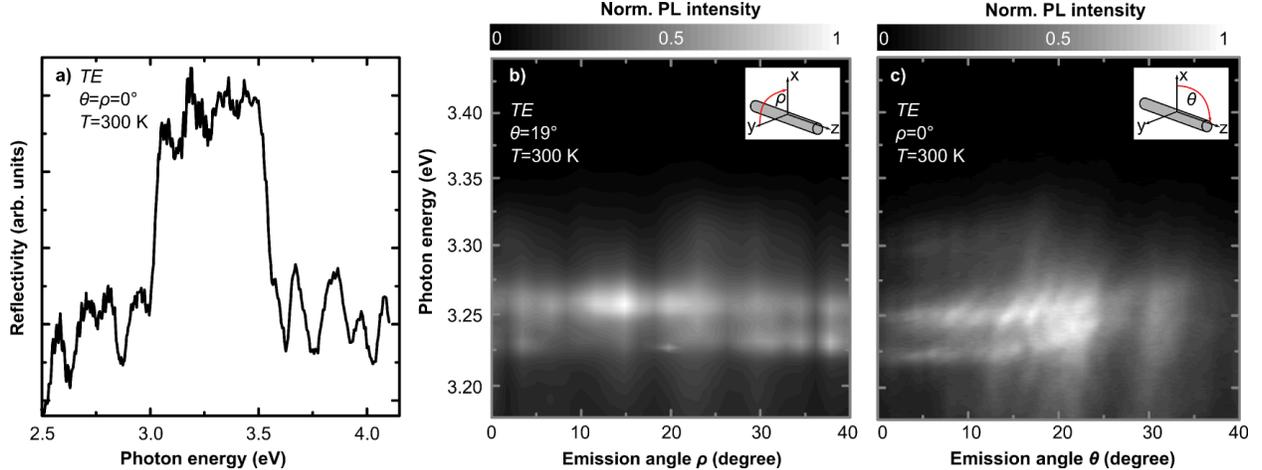}}
 \caption{ (a) Reflectivity spectrum for normal incidence. (b) and (c) angular resolved PL spectra according to the angle definitions given in the sketches. All data were recorded at room temperature.}
\label{fig:confine}    
\end{figure*}

Micro reflectivity ($\upmu$R) measurements at room temperature under normal incidence to
the cavity axis show that the Bragg stop band (BSB) is centered around $3.3\ \mathrm{eV}$ having a spectral width of $0.5\ \mathrm{eV}$ (Fig. \ref{fig:confine} a). A direct consequence of this lateral confinement is the quantization of the wave vector component in the plane perpendicular to the cavity axis as shown in the angular resolved PL spectra in Fig. \ref{fig:confine} revealing a flat mode dispersion. In contrast to that for changing the detection angle $\theta$ with respect to the cavity axis these modes shift towards higher energies with increasing $\theta$. This proves the lateral confinement and shows that optical modes can propagate only along the cavity axis.
      
It has to be mentionend that for both, TE (electrical field perpendicular to the wire axis) and TM (electrical field parallel to the wire axis) polarization, different eigenmodes are detectable in angular resolved PL measurements. But because of the fact that all TM modes are at least one order of magnitude weaker in emission intensity we will focus only on TE modes in the experimental section of this paper. As TM modes couple mainly to the high energetic C excitons they are not preferentially occupied in PL experiments which could explain their low emission intensity.   

\subsection{Mode structure}
\begin{figure*}[htbp]
\centering 
\resizebox{1\textwidth}{!}{%
  \includegraphics{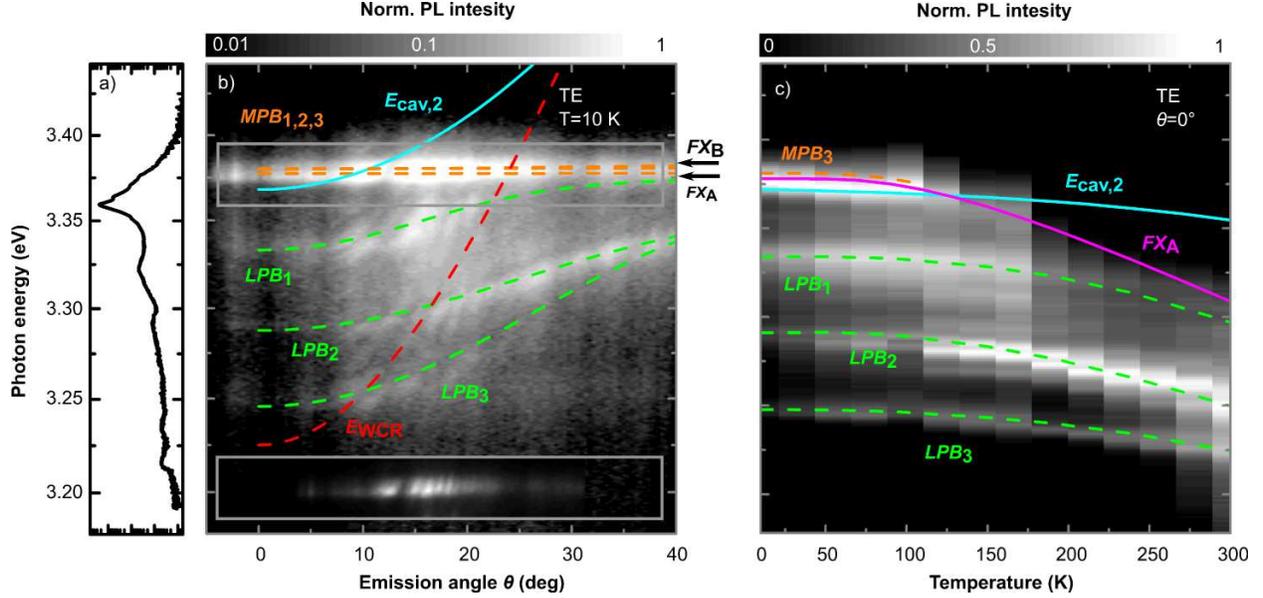}}
 \caption{Photoluminescence measurements: (a) TE polarized PL spectrum from an uncoated ZnO nanowire at $T=10\ \mathrm{K}$. (b) Angular resolved PL image at $T=10\ \mathrm{K}$ from the complete cavity. Lower (green lines) and middle (orange lines) polariton branches are modeled by the coupling of bare cavity modes (e.g. blue line) with A and B excitons. The red dashed line follows one of the weakly coupled cavity modes. The inset in (b) shows the spectral range of the excitons in a linear gray scale for a better visualization of the weakly coupled modes which cross the exciton energies. (c) Measured temperature evolution of  $\theta=0^\circ$ PL spectra for TE polarization including the thermal shift of the A exciton (pink line) together with one uncoupled cavity mode and the modeled LPBs. For each temperature the spectrum was normalized to the maximum intensity for a better visibility.} 
\label{fig:disp}    
\end{figure*}

We performed temperature dependent and angular resolved PL measurements near the center position of the cavity where its diameter is largest.
At $T=10\ \mathrm{K}$ for TE polarization, as shown in Fig.~\ref{fig:disp} b), we can clearly identify three different mode families by their angular dispersion properties. On the one hand, there are modes with a steep dispersion (e.g. red dashed line) clearly crossing the A and B excitonic ground states, as highlighted in the inset in Fig.~\ref{fig:disp} b). These modes have a energy spacing of $\Delta E_{\mathrm{WCR}}\approx 10\ \mathrm{meV}$. On the other hand, there exist modes (for $\theta=0^\circ$ at $3.25\ \mathrm{eV}$, $3.29\ \mathrm{eV}$ and $3.335\ \mathrm{eV}$; green dashed lines) whose dispersions tend to flatten when approaching the excitonic transitions at higher momenta/angles. The mode with lowest energy cannot be traced beyond to angles larger than $\theta\approx 30^\circ$ for reasons unknown.
Finally a bright emission band with a nearly flat angular dispersion is observed around $3.38\ \mathrm{eV}$, corresponding to the spectral region of the A and B exciton ground state resonances at $T=10\ \mathrm{K}$ (orange dashed lines).

The dispersion behaviour shows immediatly, that the mode family mentioned first is in the weak coupling regime (WCR) whereas the latter ones can be attributed to be in the strong coupling regime (SCR).
The simultaneous appearance of modes in WCR and SCR results from the twodimensional design of the cavity, as shown in Sec.~\ref{sec:fdtd}. It leads on the one hand to the formation of resonant modes existing only in the DBR which have no spatial overlap with the ZnO core and, on the other hand, to modes having spatial overlap with the core wire.

By applying a coupled oscillator model for the modes which are supposed to be in the SCR, including the A and B excitonic ground state energies $FX_\mathrm{A}$ and $FX_\mathrm{B}$ as well as three uncoupled photonic modes $E_\mathrm{cav,i}$ ($\mathrm{i}=1,2,3$), we can reproduce the experimental results very well as shown in Fig.~\ref{fig:disp} b). The Hamiltonian for coupling of the excitons to each of the cavity modes $E_\mathrm{cav,i}$ with the corresponding coupling constant $V_\mathrm{i}$ for this case reads~\cite{Richter2015}:
\begin{equation}
H=\left(
\begin{matrix} FX_\mathrm{A} & 0 & V_\mathrm{i}  \\ 0 & FX_\mathrm{B} & V_\mathrm{i}  \\ V_\mathrm{i} & V_\mathrm{i} & E_\mathrm{cav,i} \end{matrix}
\right)
\label{eq:hamil}
\end{equation}
The C exciton is not included in this coupling Hamiltonian as the oscillator strength for this exciton in TE configuration is negligibly small. We assume for A and B excitons the same coupling constant $V_\mathrm{i}$ because of their almost identical oscillator strength. 
As for different cavity modes it is possible to have a different spatial overlap with the core wire excitons, as shown by the simulations in Fig.~\ref{fig:simulation}, we consider for each mode i a different coupling constant. The applied model for the temperature and angular dispersions of the bare cavity modes $E_\mathrm{cav,i}$ are based on ellipsometrical studies from a planar cavity~\cite{Sturm2009} where the excitonic contribution in the DF has been computationally removed. This inhibits some inevitable uncertainties as the bare cavity angular mode dispersion also depends on the modal overlap with the core wire. The reason for the steeper dispersion of the experimental observed modes in WCR compared to the simulated dispersion of  the bare cavity modes $E_\mathrm{cav,i}$ is again found in the overlap of these modes with the ZnO core wire. The dispersion of bare cavity modes is less steep as the mode overlap with the core wire without excitons still leads to a higher effective refractive index seen by these modes compared to the modes in WCR which are located in DBR region only.       
The three eigenvalues of Hamiltonian (\ref{eq:hamil}) are in general known as upper, middle and lower polariton branch (short: UPB, MPB and LPB).

\begin{figure}
\centering 
\resizebox{0.5\textwidth}{!}{%
  \includegraphics{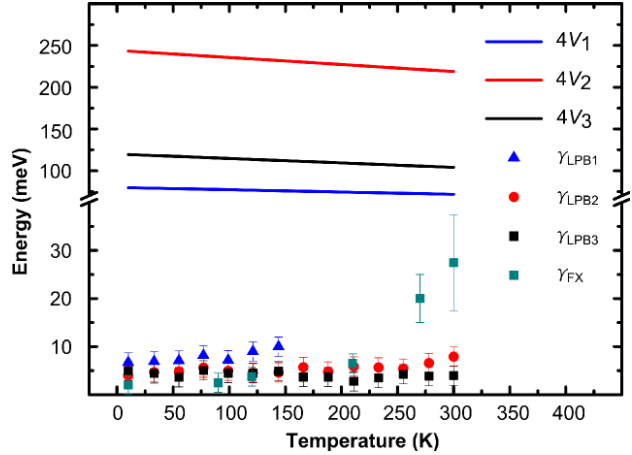}}
 \caption{Measured peak broadenings of the different LPBs and the bulk A exciton (symbols). Colored lines correspond to the fitted coupling constants $V_{\mathrm{i}}$ showing the condition for strong coupling, $\lvert \gamma_{\mathrm{FX}}-\gamma_{\mathrm{c,i}}  \rvert<4V_{\mathrm{i}}$, is fulfilled for all observable LPBs.} 
\label{fig:scr}    
\end{figure}

The three energetically lowest modes (marked green in Fig. \ref{fig:disp} b) are attributed to be the LPBs of the system. The corresponding coupling constants are obtained from modelling: $V_\mathrm{1}=20\ \mathrm{meV}$, $V_\mathrm{2}=61\ \mathrm{meV}$ and $V_\mathrm{3}=39\ \mathrm{meV}$ at $T=10\ \mathrm{K}$. This leads to detuning values $\Delta_\mathrm{1}=-27\ \mathrm{meV}$,  $\Delta_\mathrm{2}=-9\ \mathrm{meV}$ and  $\Delta_\mathrm{3}=-104\ \mathrm{meV}$, with $\Delta_{\mathrm{i}}=E_\mathrm{cav,i}-FX_\mathrm{A}$ being the detuning between the A exciton and the uncoupled cavity mode ground state.       
Furthermore the temperature evolution of these LPBs up to room temperature, see Fig. \ref{fig:disp} c), can be reproduced with the coupled oscillator model (\ref{eq:hamil}) assuming a linear reduction of the coupling strengths $V_\mathrm{i}$ of 10 percent between $10\ \mathrm{K}$ and room temperature similar to planar ZnO based cavities~\cite{Sturm2009}. 
In Fig. \ref{fig:scr} the measured spectral broadenings $\gamma$ (HWHM) of the three different LPBs and of the bulk excitons are plotted as a function of temperature.
The spectral broadening of LPB$_{\mathrm{1}}$ increases from 7 to 10 meV in the temperature range from 10 to 150 K. This can be explained by its spectral vicinity to the excitons and the relatively small coupling constant leading to an increased reabsorption due to the excitons and a possible break up of the SCR at elevated temperatures. In contrast, spectral broadenings of $\sim 5\ \mathrm{meV}$ of the lower energy LPB$_{\mathrm{2}}$ and LPB$_{\mathrm{3}}$ are almost constant. The smallest value for the LPB broadening is found at the lowest energy  LPB$_{\mathrm{3}}$. As LPB$_{\mathrm{3}}$ is supposed to have a very small excitonic contribution one can conclude that the bare cavity mode  ($E_\mathrm{cav,3}$) has a similar linewidth $\gamma_\mathrm{LPB3}~\sim\gamma_\mathrm{cav,3}=3.5 \ \mathrm{meV}$  leading to quality factor of the cavity modes of about $Q_{\mathrm{cav}}\sim 500$. This proves that the condition for strong coupling, $\lvert \gamma_{\mathrm{FX}}-\gamma_{\mathrm{cav,i}}  \rvert<4V_{\mathrm{i}}$, is fulfilled at least for LPB$_{\mathrm{2}}$ and LPB$_{\mathrm{3}}$ up to room temperature~\cite{Savona1995}. For the modes in WCR we find broadenings of $\gamma_\mathrm{WCR}\sim 1.5 \ \mathrm{meV}$ which corresponds to a quality factor of $Q_\mathrm{WCR}\sim 1000$.

As can be seen from the temperature dependent PL measurements in Fig.~\ref{fig:disp} c) the relative occupation of the different LPB ground states changes with temperature. Below $T=100\ \mathrm{K}$ the highest energy LPB is favourably occupied which changes towards room temperature where the lower energy LPB ground states gain in occupation. This may be related to a higher phonon concentration at elevated temperatures enabling a faster polariton relaxation towards the system's ground state.

\begin{figure*}
\centering 
\resizebox{1\textwidth}{!}{%
  \includegraphics{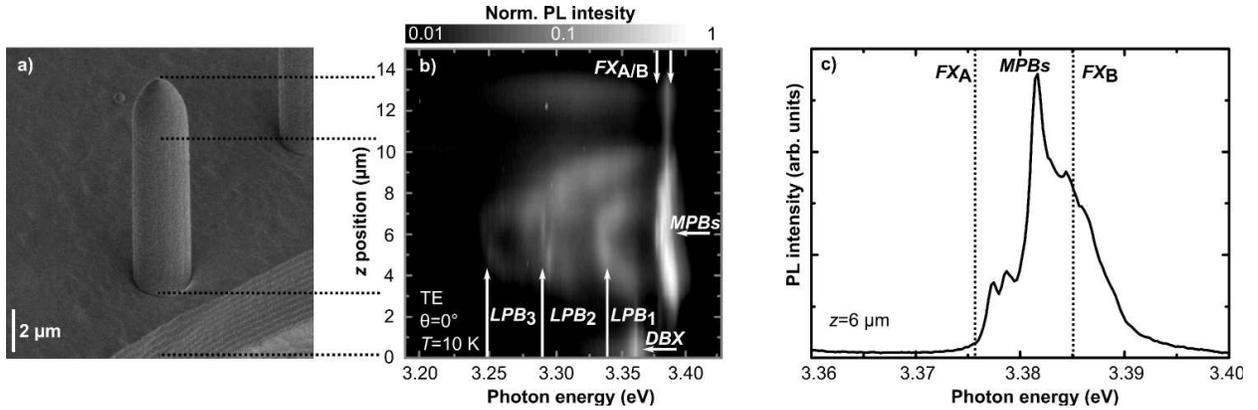}}
\caption{Scanning electron microscopy image of the nanowire cavity (a) and a confocal PL linescan (b) along the wire axis for TE polarization at $T=10\ \mathrm{K}$. DBX marks the spectral position of the defect-bound excitons from the ZnO nucleation layer. (c)  High resolution PL spectrum from the excitonic region for $\theta=0^\circ$ at the center of the cavity at $6\ \mathrm{\mu m}$.}
\label{fig:linescan}    
\end{figure*}

The emission band at $3.38\ \mathrm{eV}$ in Fig.~\ref{fig:disp} b) is attributed to be the agglomeration of all MPBs being present in the system. Their total number corresponds to the number of cavity modes $E_\mathrm{cav,i}$ being present and showing strong coupling to the ZnO excitons~\cite{Richter2015}. One may assume that the emission at $3.38\ \mathrm{eV}$ is created by defect-bound excitons, the brightest and dispersionless light sources emitting from ZnO bulk samples in low temperature PL experiments. But the comparison of the PL emission between an uncoated core wire, as shown in Fig. \ref{fig:disp} a), and the complete cavity proves that this is obviously not the case as the recombination from defect-bound excitons are located at lower energies.
A final proof for the presence of MPBs in our cavity structure is given by spatially resolved PL linescans along the wire axis, as shown in Fig. \ref{fig:linescan} b), where the cavity thickness dispersion of the MPBs is probed. Here we can resolve the small energy dispersion of the MPBs towards higher energies by approaching the cavity ends where the core wire diameter is smallest.
Furthermore in this spectral linescan the PL emission from the ZnO nucleation layer at the bottom of the cavity is observed, which is again dominated by the emission of the defect-bound excitons. From the uppermost two microns of the cavity almost no emission is detected, as there is no ZnO core wire in this region. Furthermore a high resolution PL spectrum (Fig.~\ref{fig:linescan} c) around the $3.38\ \mathrm{eV}$ peak shows that it really is composed of different modes, whose spectral spacing is much smaller than that of the weakly coupled modes of about $10\ \mathrm{meV}$. This strongly supports the theory of multiple MPBs.
The appearance of these MPBs is a direct consequence of the high quality of the ZnO core wire material as the excitons are not remarkably broadened by structural disorder leading to a free spectral range between A and B exciton where reabsorption is negligible. This clear appearance of MPBs in luminescence experiments is unique in present day literature about ZnO-based cavities. The same holds for GaN-based cavities, which has properties comparable to ZnO. 

The emission from the MPBs vanishes above $150\ \mathrm{K}$, see Fig. \ref{fig:disp} c). This can be explained by the temperature induced broadening of the A and B excitons which exceeds the free spectral range of $9\ \mathrm{meV}$ between these excitons (see also Fig.~\ref{fig:df} c) in the Appendix). Above this critical temperature the MPBs are reabsorbed by the excitons. 

As typical for ZnO-based cavities the UPBs cannot be observed in PL measurements. This is caused by the large coupling constant in combination with a strong polariton-phonon interaction in ZnO which makes the occupation of the UPBs less probable due to fast scattering into lower energy states, and by reabsorption~\cite{Faure2008,Lin2011}.

\subsection{Nonlinear emission properties}
\begin{figure*}
\centering 
\resizebox{1\textwidth}{!}{%
  \includegraphics{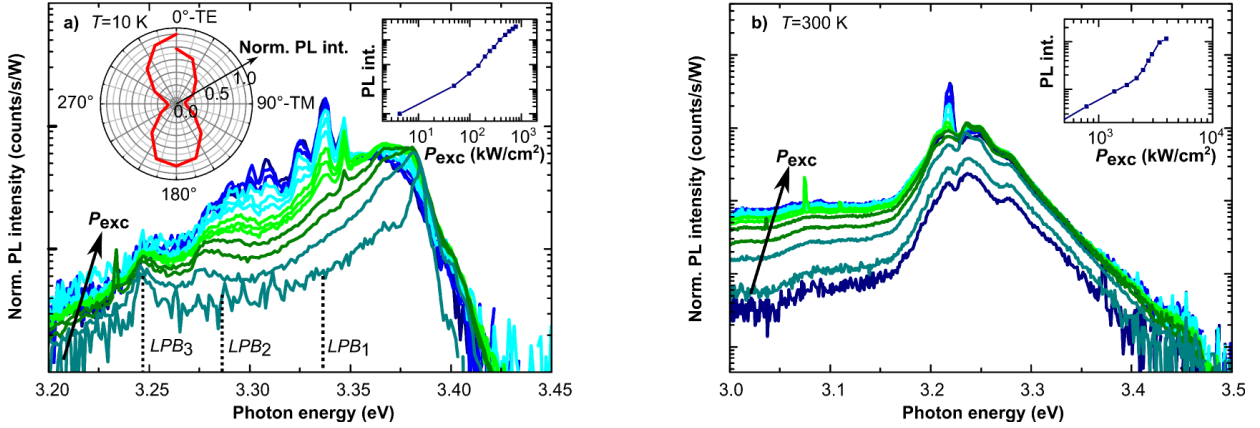}}
\caption{Excitation power dependent PL spectra for  $\theta=0^\circ$ at (a) $T=10\ \mathrm{K}$  and (b) $T=300\ \mathrm{K}$. These spectra have been normalized to the applied excitation power density $P_\mathrm{exc}$ and integration time. The left inset in (a) shows exemplarily the linear polarized emission intensity in dependence on the polarization angle. Both panels also include an inset showing the emitted PL intensity vs. excitation power density.}
\label{fig:lasing}    
\end{figure*}

Excitation dependent PL measurements at $T=10\ \mathrm{K}$ and $T=300\ \mathrm{K}$, as shown in Fig.~\ref{fig:lasing}, reveal nonlinear emission behavior. For low temperatures at a threshold power density of $P_\mathrm{th}\approx 100\ \mathrm{kW}/\mathrm{cm}^2$ a superlinear increase in emission intensity can be observed in the spectral range below the A and B excitons in combination with the appearance of narrow modes having a spectral spacing of about $10\ \mathrm{meV}$. These modes are almost purely TE polarized and extended over the full observable $\theta$-range. The degree of polarization is $\Pi=(I_\mathrm{TE}-I_\mathrm{TM})/(I_\mathrm{TE}+I_\mathrm{TM})=94\%$. The relatively high threshold power density in combination with the redshift of the underlying gain profile leads to the conclusion that the observed nonlinear phenomena can be explained with the formation of an electron-hole plasma providing gain for stimulated emission out of weakly coupled modes~\cite{Yamamoto2002}.

At room temperature single mode lasing can be observed in a cavity out of the same growing charge as the one described before. Here the threshold power density is found to be $P_\mathrm{th}\approx 2000\ \mathrm{kW}/\mathrm{cm}^2$. At both temperatures the emission out of the LPBs can still be observed at or above the threshold. This coexistance is caused by the almost Gaussian intensity distribution of the laser spot profile on the sample. In its center, where the intensity is highest, lasing sets in firstly whereas at the rim of the excitation laser spot the critical density is not reached and polaritons can still exist.

The reason why we are not observing polariton Bose-Einstein condensation is probably found in the relatively low $Q$ factors of the cavity modes in combination with a small mode overlap (see Sec.~\ref{sec:fdtd}) of the cavity modes with the ZnO core wire. The latter aspect leads to a reduced coupling constant $V$ in comparison to the bulk polariton one which should exceed several hundred meV. Furthermore the high density of weakly coupled modes may deplete the exciton population in the core wire via the Purcell effect~\cite{Purcell1946}.

\section{Conclusion}
We have shown that our approach of coating self-assembled grown ZnO nanowires with radial symmetrical distributed Bragg reflectors leads to high quality cavity structures. Here the simultaneous coupling of several photonic modes with the ZnO A and B excitons was observed with the unambiguous appearance of so-called middle polariton branches. In addition to that we simultaneously observe modes in the weak coupling regime due to a missing mode overlap with the core excitons as we have shown by means of FDTD simulations recovering the spatial mode distributions inside the cavity structure. Also the polariton modes are reproduced by these simulations. Furthermore, we observed lasing phenomena up to room temperature in our cavities. At low temperature we find multimode lasing whereas at room temperature singlemode lasing can be observed.

\section{Acknowledgements}
We thank C. Sturm for fruitful discussions and J. Lenzner for technical support. This work was funded by the Deutsche Forschungsgemeinschaft within Forschergruppe 1616 \linebreak (SCHM2710/2).

\section{Appendix}

\begin{figure*}
\centering 
\resizebox{1\textwidth}{!}{%
 \includegraphics{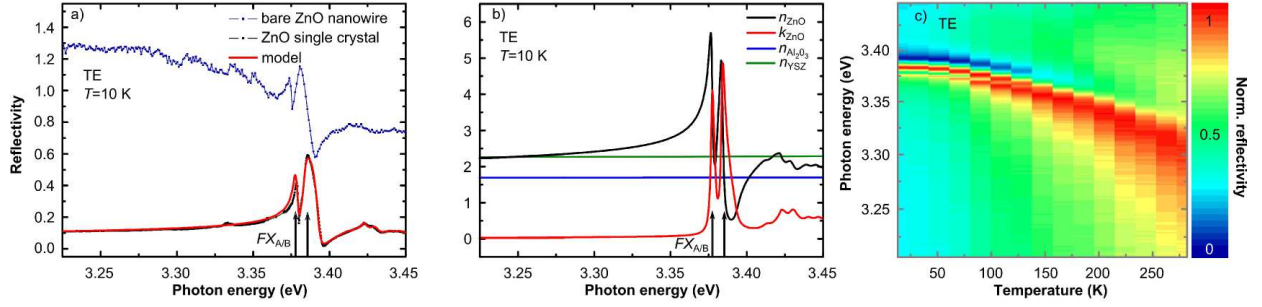}}
\caption{(a) Reflectivity spectrum (black line and squares) under normal incidence of an $a$-plane cut taken from a ZnO single crystal at $T=10\ \mathrm{K}$ is shown exemplarily together with the fitted model (red line). The graph also includes a reflectivity spectrum (shifted towards higher reflectivities for clarity) from a bare ZnO nanowire for polarization perpendicular to the optical axis (blue line and squares). (b) Modeled values for the refractive index $n$ (red line) and the extinction coefficient $k$ (black line) are plotted vs. photon energy together with the refractive indices of the transparent DBR materials (green and blue lines). (c) Temperature dependent reflectivity spectra from the ZnO single crystal normalized to their corresponding maximum for a better visibility of the temperature evolution.}
\label{fig:df}    
\end{figure*}

In order to determine the ZnO exciton energies and broadenings in dependence on the sample temperature we performed polarization resolved microreflectivity measurements at an $a$-plane cut taken from a high-quality ZnO bulk single crystal (Fig.~\ref{fig:df}). Furthermore we measured the reflectivity from a bare ZnO nanowire (Fig.~\ref{fig:df}~a) for polarization perpendicular (TE) to the wire axis. Here an exact reflectivity measurement cannot be obtained as the wire diameter is considerably smaller than the probe spot. Also the reflectivity probe spot diameter is slightly dependent on the wavelength of the probe light which results in a spectral reflectivity which is differently weighted compared to the results from the single crystal. But nevertheless the excitonic  features are reproduced indicating similar optical properties of the nanowire and the bulk single crystal. Thus, using optical data obtained from the bulk single crystal for modeling the dielectric function (DF) of the ZnO nanowire is warrantable.

 A model DF~\cite{Bundesmann2008,Schmidt-Grund2011} has been applied to fit the experimental data of the single crystal. This model DF was also used as input parameters for the FDTD simulations. The refractive indices of the DBR layers were deduced by spectroscopic ellipsometry measurements from planar thin film samples.  

\end{document}